\documentclass[twocolumn,aps,prl,superscriptaddress]{revtex4}
\usepackage{graphicx}
\usepackage{amssymb}


\begin{document}

\title{Recovery of Hidden Interference in Mott Insulators}
\author{L. Tian}
\affiliation{Department of Applied Physics and E. L. Ginzton Laboratory, Stanford University, Stanford, CA 94305}
\author{F. Fujiwara}
\affiliation{National Institute of Informatics, 2-1-2 Hitotsubashi, Chiyoda-ku, Tokyo 101-8430, Japan}
\author{T. Byrnes}
\affiliation{National Institute of Informatics, 2-1-2 Hitotsubashi, Chiyoda-ku, Tokyo 101-8430, Japan}
\affiliation{Institute of Industrial Science, University of Tokyo, 4-6-1 Komaba, Meguro-ku, Tokyo 153-8505, Japan}
\author{Y. Yamamoto}
\affiliation{Department of Applied Physics and E. L. Ginzton Laboratory, Stanford University, Stanford, CA 94305}
\affiliation{National Institute of Informatics, 2-1-2 Hitotsubashi, Chiyoda-ku, Tokyo 101-8430, Japan}

\date{\today}

\begin{abstract}
Particle statistics plays a crucial role  in a strongly interacting quantum many-body system. 
Here, we study the Hubbard model for distinguishable particles at unit filling. 
Starting from the superfluid-like state in the strong tunneling limit and gradually reducing the 
tunneling so that the on-site repulsive interaction dominates, the state ends up in a symmetric 
superposition of Mott insulator states.  This result can be experimentally confirmed by the recovery of 
interference patterns in the density correlation functions.  We also show that this state is a 
maximally entangled state, in contrast to the standard picture.
\end{abstract}
\maketitle

%introduction
%%%%
The Bose-Hubbard model is a well studied model of spinless interacting bosons  \cite{FisherPRB1989}.    
In this model, the bosons interact with on-site repulsive interactions and can tunnel between 
adjacent lattice sites. A quantum phase transition occurs between a superfluid state in the 
strong tunneling limit and a Mott insulator state in strong interaction limit, as a result of 
the competition between the tunneling and the interaction \cite{SachdevBook1994}.  Recently, this 
model has been experimentally realized using ultra-cold atoms in optical lattices and a phase 
transition has been observed by examining the diffraction patterns of the atoms \cite{GreinerNature2002}.   

The Mott insulator state was experimentally verified by the detection of a single Gaussian 
distribution in the diffraction pattern. Particles in this state are seemingly  localized in each lattice site with 
vanishing spatial coherence over other lattice sites.  In a more recent experiment, the measured 
spatial correlation function reveals higher-order coherence in this
state \cite{FollingNature2005,GreinerPRL2005}.  However, because of the symmetrization postulate for 
identical bosons, it is  impossible to determine whether the observed interference is a result of the 
trivial coherence due to the symmetrization postulate or is a result of the intrinsic coherence 
generated by particle-particle interaction.   

In this paper,  we study a Hubbard model for distinguishable particles (as opposed to identical
bosons or fermions) to clarify the hidden 
coherence and entanglement in  the Mott insulator state. 
We will show that quantum interference can be found for distinguishable particles, which demonstrates 
the  (truly) coherent nature of the Mott insulator state, in which each particle not only occupies 
all sites in a linear superposition state but also has correlations with other particles. The 
theoretical prediction of the recovery of the interference patterns can be experimentally tested 
by a relatively small scale experimental system. Using exact diagonalization methods, we derive ground state properties of the Hubbard model 
for distinguishable particles. We show that the ground state of this model is a fully symmetric 
state, in complete analogy to its bosonic counterpart. The distinguishability causes a large 
degeneracy in the Mott insulator states in the limit of zero tunneling. At small but finite 
tunneling, the ground state can be approximated as a symmetric superposition of such degenerate 
Mott insulator states.  Quantum interference \cite{CKHongPRL1987,GritsevNPhys2006} appears in the 
correlation function for coincidence detection of particles because of  this (non-trivial) 
superposition.  A quantum many-body system with distinguishable particles can also be very useful 
in studying entanglement  in the quantum phase transition, as it avoids the subtle issue associated 
with entanglement and quantum indistinguishability.  We compare the entanglement in the distinguishable 
particle Hubbard model and that in the standard Bose-Hubbard model.  We adopt an operational definition of the 
entanglement following Ref.  \cite{WisemanPRL2003}, where entanglement is the correlation between 
modes (not particles) in subsystems.  Very different results are obtained for the two models.  
%%%%

% distinguishable hubbard model
%%%%
Consider a Hubbard model for distinguishable particles in a one-dimensional periodic lattice.  
The particles can tunnel between nearest neighbor sites with tunneling matrix element 
$t_0$ which is the same for all species of distinguishable particle (henceforth called an ``isotope''). 
Particles located on the same lattice site interact with each other with a repulsive 
interaction $U_0$ for all isotopes. In order to directly compare with the Bose-Hubbard model, we define an operator-based Hamiltonian for the distinguishable model.   The Hamiltonian
can be written as:
\begin{equation}
H=-t_{0}\sum_{\alpha,i}\left ( c_{\alpha,i}^{\dagger}c_{\alpha,i+1}+h.c.\right )+
\frac{U_{0}}{2}\sum_{i}n_{i}(n_{i}-1)\label{eq:Hc}
\end{equation}
where $n_{i}=\sum_{\alpha}n_{\alpha i}$ is the total number of particles on site $i$, and 
$\alpha$ labels the isotopes. The operators $c_{\alpha,i}$ and 
$c_{\alpha,i}^\dagger$ are the annihilation and creation operators for isotope $\alpha$, 
and satisfy the commutation relations 
$[c_{\alpha,i},c_{\beta,j}^\dagger]= \delta_{\alpha \beta} \delta_{i j}$ 
and $[c_{\alpha,i},c_{\beta,j}]=0$.   When the index $\alpha$ is omitted,  the model becomes the standard Bose-Hubbard model.  

\begin{figure}
\includegraphics[bb=54bp 240bp 576bp 666bp,clip,width=7.5cm]{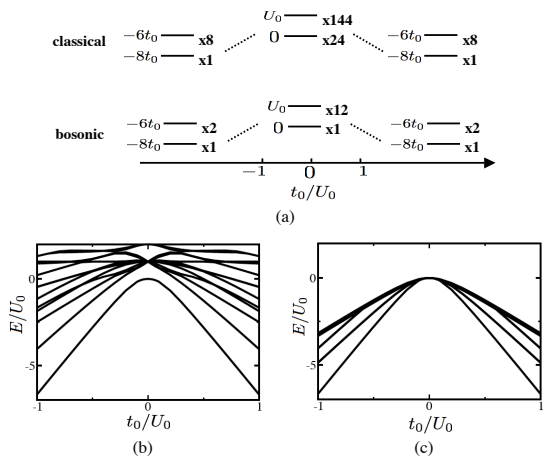}
\caption{Ground state properties with $N=4$. (a) Schematic energy structure of the ground states and the first excited states.  Energy levels are shown as solid lines with its energy labeled  on the left hand side and its degeneracy labeled on the right hand side. Energy spectrum versus $t_0/U_0$ are plotted for the lowest $15$ states in  (b) the Bose-Hubbard model  and (c) the Hubbard model for distinguishable particles.  }\label{fig:1}
\end{figure}

We are interested in the case of unit filling for a finite lattice with a total of $N$ sites, 
where the number of particles equals the number of lattice sites. 
We also only consider the case where there is one particle per isotope
$N_\alpha = \sum_i c_{\alpha,i}^{\dagger}c_{\alpha,i}=1$. 
At $t_{0}\gg U_{0}$, where the 
on-site interaction is negligible, each particle occupies the lowest single particle mode of 
the tunneling Hamiltonian. The  ground state is a non-degenerate state with wave function
\begin{equation}
|\psi_{SF} \rangle=\prod_{\alpha=1}^{N}c_{\alpha,k=0}^{\dagger}|0\rangle \label{eq:bec}
\end{equation}
where $c_{\alpha,k}^{\dagger}=\sum_{n}e^{ikn} c_{\alpha,n}^{\dagger}/\sqrt{N}$ is the creation operator of the 
a single particle mode of momentum $ k$ for isotope $\alpha$.  This state, with an energy of $-2N t_0$, corresponds  
to the superfluid limit of the Bose-Hubbard model and is symmetric under exchange of any two particles. 
In the first excited state, one particle occupies the second single particle mode. In an $N=4$ lattice, 
the excited states are 8-fold degenerate with an energy $2t_0$ above the ground state as shown in 
Fig.~\ref{fig:1} (a).   At $t_{0}=0$, the particles avoid each other because of the repulsive interaction 
and localize  at different lattice sites. The ground states are degenerate states with $N!$-fold 
degeneracy because of the distinguishability of the particles. The wave functions have the form
$|\psi_{MI}^{P}\rangle= \prod_{\alpha=1}^{N} c_{\alpha,i=P(\alpha)}^{\dagger} |0\rangle$
where the $P(\alpha)$ is a permutation of the list $\{ 1,\cdots, N\}$. These states correspond 
to the Mott insulator state of the Bose-Hubbard model. With $N=4$, there are $24$ degenerate states, 
which are shown in Fig.~\ref{fig:1} (a) together with the excited states.

Exact matrix diagonalization is used to 
study the ground state properties of the above model.  We calculate the eigenstate spectrum, 
the eigenfunctions, and the fringe visibility for a range of $t_0/U_0$. 
The energy spectrum with $N=4$ particles is plotted in Fig. \ref{fig:1} (c).  We find that 
the ground state energy and wave functions of the distinguishable particles are in exact 
agreement with that of the bosons (Fig. \ref{fig:1} (b)) everywhere in the parameter regime, 
except at zero tunneling.  At zero tunneling, the statistics of the particles causes high degeneracy in the ground states.  
At $t_0\ne 0$ but $t_0\ll U_0$, the ground state can be 
approximated as the symmetric superposition of the Mott insulator states:  
\begin{equation}
|\psi_{MI} \rangle=\frac{1}{\sqrt{N!}}\sum_{P}\prod_{\alpha=1}^{N} c_{\alpha,i=P(\alpha)}^{\dagger} |0\rangle 
. \label{eq:psi_sym}
\end{equation}  
The symmetry in the Mott insulator state at $t_0\ll U_0$ is hence not a trivial consequence of the 
symmetrization postulate. Even though the gap between this ground state and the excited states is 
$\sim t_0^2/U_0$, very small at this limit, it can be prepared experimentally by an adiabatic approach.  
Starting from the superfluid-like state at strong tunneling limit and slowly turning off the tunneling 
to $t_0\ll U_0$, the final state becomes the symmetric linear superposition state before it is 
thermalized by the environment.

\begin{figure}
\includegraphics[bb=90bp 420bp 540bp 660bp,clip,width=7cm,keepaspectratio]{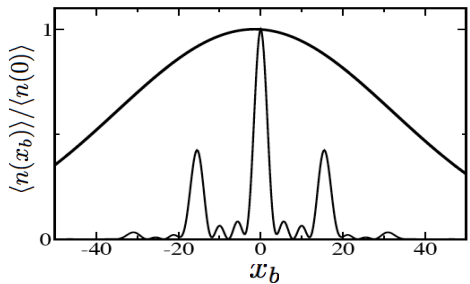}
\caption{Particle detection (single particle detection probability) versus position $x_b$ at time $t$ for the Mott insulator state (thick solid curve) and for the superfluid-like state (thin solid line) with $N=4$.} \label{fig:2}
\end{figure}

The visibility of the fringes in the single particle diffraction pattern can be viewed as an order 
parameter for a finite lattice model and is defined as  
\begin{equation}\displaystyle
V=\frac{\max{\{\langle n_{k}\rangle \}}-\min{\{\langle n_{k}\rangle \}} }{ \max{\{\langle n_{k}\rangle \}}+
\min{\{\langle n_{k}\rangle \}}} \label{eq:Vnk}
\end{equation}
with $n_{k}=\sum_{\alpha}c_{\alpha,k}^{\dagger}c_{\alpha,k}$ being the number operator in $k$-space and 
$k=2\pi n/N$ for $n=0,\cdots, N-1$.  It is an important parameter for studying quantum phase 
transition \cite{ScarolaPRA2006}.   As discussed above, at $t_0\gg U_{0}$, 
$\max{\{ \langle  n_{k}\rangle \}}=N$ at $k=0$ and $\min{ \{ \langle  n_{k}\rangle \}}=0$ for all 
other $k$, we have $V=1$; while at $t_{0}\ll U_{0}$, $\langle  n_{k} \rangle =1$ for all $k$, we 
have $V=0$.  Between these two limiting cases, the visibility fringe monotonically increases 
from $0$ to $1$, clearly demonstrating the change in nature of the ground state between the Mott 
insulator to the superfluid-like states  (see also Fig. ~\ref{fig:2}) for the distinguishable particles.
%%%%

% interference
%%%%
The detection of the correlation functions is a more powerful tool in studying the nature of a quantum 
many-body system than single particle detection \cite{CKHongPRL1987,GritsevNPhys2006}.  In this process, the particles in their ground state
are released from the lattice and the spatial  density  correlation is measured after a sufficiently 
long time.  To study this process quantitatively, we treat each lattice site as an harmonic potential, 
but with an energy barrier between adjacent sites that is half of the harmonic potential: 
$V_0=m\omega^{2}a_{0}^{2}/16$, where $a_0$ is the size of the unit cell in the lattice, 
$m$ is the mass of the isotopes, and $\omega$ is the frequency of the harmonic potential. 
We describe the wave function of the localized state at site $i$ as 
\begin{equation}
\langle x|c_{\alpha,i}^{\dagger}|0\rangle=\phi(x-R_{i})\label{eq:phi_x}
\end{equation} 
with  $\phi(x)=\frac{1}{^{4}\sqrt{\pi\sigma}}e^{-x^{2}/2\sigma}$ in the Gaussian approximation. 
The position uncertainty of this state is $\sigma=\hbar/m\omega$. Estimating the tunnelling amplitude
using $t_{0}\sim \hbar\omega e^{-V_{0}/\hbar\omega}$ \cite{LandauBook1977}, the tunneling matrix element can be expressed as  
$ t_{0}=(\hbar^2 /m\sigma )e^{-a_{0}^{2}/16\sigma }$ in terms of $\sigma$.  
Below, we assume $8\hbar^2/ma_{0}^{2}=5U_{0}$ in the calculation.  After being suddenly released from 
the lattice, the particles evolve freely with the transformation 
$U(t)=\exp{(-i\sum_\alpha\frac{p_\alpha^{2}}{2m\hbar}t )}$ at time $t$. Here, $p_\alpha$ is the 
momentum operator of particle $\alpha$. In the coordinate basis, we have 
\begin{equation}
\langle x_\alpha|U(t)c_{\alpha,i}^{\dagger}|0\rangle=\sqrt{\frac{\pi}{im/2t\hbar+1/\sigma}}
e^{-\frac{(x_\alpha-R_{i})^{2}}{\sigma-i2t\hbar/m}}\label{eq:Uxt}
\end{equation}
for particle $\alpha$. 

The density operator at position $x_b$ can be written as 
$n(x_b)=\sum_\alpha |x_{\alpha}=x_b \rangle \langle x_{\alpha}=x_b|$, including all isotopes.  
Using the wave functions derived above, the density of the particles  at time $t$, also called the 
first order correlation,  can be calculated as $ \langle\psi |U^{\dagger}(t) n(x_{b}) U(t)|\psi \rangle$. 
In Fig.~\ref{fig:2}, we plot the density for both the Mott insulator state at $t_0 \ll U_0$ and the 
superfluid-like state at $t_0\gg U_0$.  For the Mott insulator state, the density shows a smooth 
distribution with a Gaussian profile resulting from the initial Gaussian state in Eq.~(\ref{eq:phi_x}). 
Note that an incoherent product state of localized states on all lattice sites has the same density 
distribution as plotted in the figure.  For the superfluid-like state, an interference pattern appears with three major peaks corresponding to particles traveling with wave vectors $k=0,\pm 2\pi/a_0$ respectively.  This result agrees with the analysis of the visibility fringe defined in Eq.~(\ref{eq:Vnk}) for large $N$. 

\begin{figure}
\includegraphics[bb=90bp 198bp 540bp 660bp,clip,width=7cm]{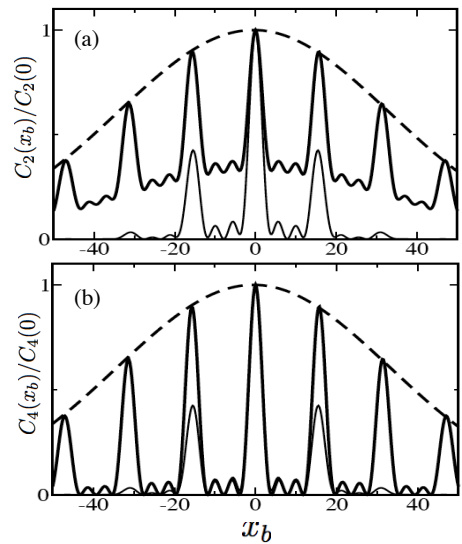}
\caption{(a) Interference in the normalized second order correlation function and (b) the normalized 
fourth order correlation function (lower plot) with $N=4$ particles.  Dashed curves:  
individual Mott insulator state; thick solid curves: symmetric superposition of the Mott 
insulator states in Eq.~(\ref{eq:psi_sym});  and thin solid curves:   superfluid-like state. }\label{fig:3}
\end{figure}

Higher order correlation functions for the coincidence detection of particles can be calculated 
similarly. We plot the second order correlation function 
\begin{equation} 
C_{2}(x_b)=\langle\psi |U^{\dagger}(t)n(0)n(x_{b})U(t)|\psi \rangle \label{eq:corr2}
\end{equation} 
in Fig.~\ref{fig:3} (a) and the fourth order correlation function 
\begin{equation} C_{4}(x_b)=\langle\psi |U^{\dagger}(t)n(0)n(x_{b})n(0)n(0)U(t)|\psi \rangle \label{eq:corr4}
\end{equation}
in  Fig.~\ref{fig:3} (b).   In the plots, the correlation functions for each (fictitious) Mott insulator state show the same Gaussian profile as their density distribution.  The correlation functions for the superfluid-like state  show the same interference pattern as their density distribution as expected (each particle is independent and uncorrelated).   In sharp contrast, interference patterns can now be seen in the higher order correlation functions for the symmetric superposition of the Mott insulator state in Eq.~(\ref{eq:psi_sym}).  These correlation functions include linear superposition of terms given by Eq. (\ref{eq:Uxt}) for all the permutations $P$. Quantum interference appears as a result of the coherent superposition and particle-particle correlation. It can be shown that for $N$ particles, the interference pattern can be found  from the second to the $N$-th order correlation function.  

In order to understand the above result, let us consider the simplest case of $N=2$. The Mott insulator 
state in this case is 
$ | \psi_{12}\rangle=\frac{1}{\sqrt{2}}(|R\rangle_1 |L\rangle_2+|L\rangle_1 |R\rangle_2)$. 
With $|\pm\rangle_{1,2}=\frac{1}{\sqrt{2}}(|L\rangle\pm |R\rangle)_{1,2}$, 
$|\psi_{12} \rangle =\frac{1}{\sqrt{2}}(|+\rangle_1 |+\rangle_2-|-\rangle_1 |-\rangle_2)$. 
If we ignore particle $2$, the state of particle $1$ is a mixed state,
$\rho_1=\frac{1}{2}(|L\rangle_1\langle L| + |R\rangle_1\langle R| )$, and there is no interference.  
However, if we project particle $2$ to $|+\rangle_2$  by detecting it, the resulting state of 
particle $1$ remains at $\frac{1}{\sqrt{2}}(|L\rangle+|R\rangle)_1$, and there is interference. 
This is a standard situation of a ``quantum eraser''. This result clarifies the origin of the 
interference pattern (and hence coherence) in the Mott insulator state. As we are dealing with 
distinguishable particles, it is clear here that the interference originates from the particle 
interactions, not due to the symmetrization postulate.

The above results are for small number of particles. For large $N$, it can be shown that 
$ n(x_b)\propto n_k$  with $x_b=\hbar k t/m$ after a releasing time $t$.  
The correlation functions can then be derived analytically at $t_{0}\ll U_{0}$ and $t_0\gg U_{0}$ using 
the wavefunctions (\ref{eq:bec}) and (\ref{eq:psi_sym}). We 
find that 
\begin{equation} 
C_{2}(x_b)|_{\psi_{SF}} \propto 
\langle n_{k}n_{k^{\prime}} \rangle = \delta_{k,0}\delta_{k^{\prime},0}N(N-1)
\label{eq:t0N}
\end{equation} 
at $t_0\gg U_0$. The $\delta$-function at $k=0$ produces the major interference peaks corresponding to 
$x=\hbar G t /m $ with $G=0, \pm 2\pi /a_0$ in Fig.~\ref{fig:2} and Fig.~\ref{fig:3}. 
At $t_0\ll U_0$, for (fictitious) individual Mott insulator states, we have  
$\langle n_{k}n_{k^{\prime} } \rangle=(N-1)/N$, which is a flat distribution over the $k$-space; 
for the symmetric superposition of Mott insulator states,  we have 
\begin{equation} 
C_{2}(x_b)|_{\psi_{SF}} \propto\langle n_{k}n_{k^{\prime}} \rangle \frac{1}{N}(N-2+\delta_{k,k^{\prime}}N),
\label{eq:U0N}
\end{equation} 
agreeing with that of (indistinguishable) bosons.   Similar results can be calculated for  higher order 
correlation functions. 
For example, we have $\langle n_{k}n_{k^{\prime}}n_{k^{\prime\prime}} 
\rangle=1/4+2\delta_{k,k^{\prime}}\delta_{k^{\prime},k^{\prime\prime}}$ for the symmetric state; 
while  $\langle n_{k}n_{k^{\prime}}n_{k^{\prime\prime}} \rangle=3/8$ for 
individual Mott insulator states.  Therefore, the interference can be observed in the correlation functions at large $N$ with reduced visibility.
%%%%

% entanglement
%%%%
\begin{figure}
\includegraphics[bb=72bp 240bp 576bp 600bp,clip,width=7.5cm]{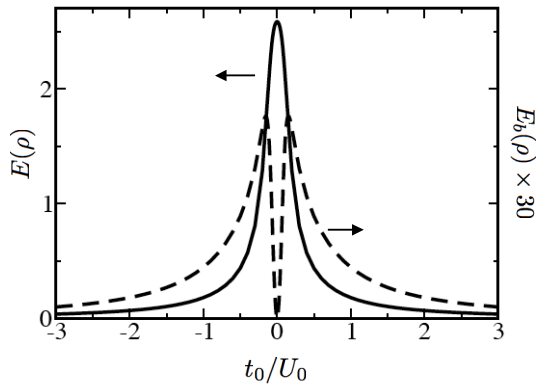}
\caption{Entanglement for distinguishable particles and identical bosons versus $t_0/U_0$ for $ N= 4$.  
Solid curve: $E ( \rho) $ for distinguishable particles. 
Dashed curve: $E_{b} (\rho) $ for bosons. }\label{fig:4}
\end{figure}
Finally, we study the entanglement in the ground states of both distinguishable particles and 
identical bosons. 
The statistics of the particles affects the entanglement in a very different way.  For 
distinguishable particles, the entanglement can be characterized by calculating the entropy $E(\rho)$ of a 
subsystem containing half the number of particles after tracing off the other half. The superfluid-like 
state at the strong tunneling limit is a separable state with $E(\rho)=0$. Starting from this state and 
gradually reducing the tunneling to zero, we obtain the symmetric superposition of the Mott insulator 
states which is a maximally entangled state with $E(\rho)\approx 2.5$, as is shown in Fig.~\ref{fig:4}. Here, entanglement is  generated because the interaction does not commute with the single particle Hamiltonian.
 
For identical bosons,  the study of entanglement is complicated by the fact that the particles are identical and the state space is much smaller than that for distinguishable particles. We use an operational definition for entanglement presented in Ref. \cite{WisemanPRL2003}. The entanglement can be described as the correlation between two subsystems that can be manipulated by local operations, with the definition
\begin{equation} E_{b}(\rho)=\sum_{n=0}^{N}p_{n}E(\prod_{n}|\psi_{g}\rangle\langle\psi_{g}|\prod_{n})\label{E2b}\end{equation}
where  $\prod_{n}$ projects the state to a subspace with $n$ particles in the left two sites.  In Fig.~\ref{fig:4}, we plot the entanglement for bosons. Instead of a monotonic increase with the interaction  $U_0$ as for the distinguishable particles, the entanglement for identical bosons $E_{b}$ reaches a peak  at $t_0/U_0\approx 0.16$ and decreases to zero at $t_0\ll U_0$ where the projected state in the subspace is a pure state with no correlation with the state in the other subsystem and $E_b=E(c_1^\dagger c_2^\dagger |0\rangle\langle 0| c_1 c_2)= 0 $.

To conclude, we studied the ground state properties of the  Hubbard model for distinguishable particles 
by examining correlation functions and entanglement properties.   In the strong interaction limit 
$t_0\ll U_0$,  
the symmetric superposition of the Mott insulator states becomes the ground state and hence the 
quantum interference in the correlation functions  and entanglement between the particles is found. 
Our results suggest that a better understanding of the role of quantum statistics in quantum phase 
transitions can be gained by studying distinguishable particles.  Using a relatively small system 
of cold atoms\cite{GreinerNature2002,JakschPRL1998}, trapped ions\cite{PorrasPRL2004}, and 
exciton polariton\cite{KasprzakNature2006,HDengPRL2006}, theoretical predictions presented in this 
paper should be experimentally confirmed.

This work is supported by  SORST program of Japan Science of Technology Corporation (JST), and NTT Basic Research Laboratories, and Special Coordination Funds for Promoting Science and Technology of Univ. Tokyo.  L. T.  is supported in part by Karel Urbanek Fellowship.


\begin{thebibliography}{10}
\bibitem{FisherPRB1989}
M. P. A. Fisher, P. B. Weichman, G.  Grinstein, and D. S. Fisher, Phys. Rev. B \textbf{40}, 546 (1989).
\bibitem{SachdevBook1994}
S. Sachdev, {\em Quantum Phase Transition}, Cambridge University Press (2001).
\bibitem{GreinerNature2002}
M. Greiner and \textit{et al.}, Nature (London) \textbf{415}, 39 (2002).
\bibitem{FollingNature2005} S. F\"{o}lling and \textit{et al.}, Nature (London) \textbf{434}, 481 (2005).
\bibitem{GreinerPRL2005} M. Greiner, C. A. Regal, J. T. Stewart, and D. S. Jin, Phys. Rev. Lett. \textbf{94}, 110401 (2005).
\bibitem{CKHongPRL1987}
C. K. Hong, Z. Y. Ou, and L. Mandel, Phys. Rev. Lett. \textbf{59}, 2044 (1987).
\bibitem{GritsevNPhys2006}
V. Gritsev, E. Altman, E. Demler, and A. Polkovnikov, Nature Phys. \textbf{2}, 705 (2006).
\bibitem{WisemanPRL2003}
H. M. Wiseman and J. A. Vaccaro, Phys. Rev. Lett. \textbf{91}, 097902 (2003).
\bibitem{ScarolaPRA2006}
V. W. Scarola, E. Demler,  and S. Das Sarma, Phys. Rev. A \textbf{73}, 051601(R) (2006).  
\bibitem{LandauBook1977}
L. D. Landau and E. M. Lifshitz,  {\em Quantum mechanics: non-relativistic theory}, 3d ed., Pergamon Press  (1977).
Pergamon Press (1977).
\bibitem{JakschPRL1998}
D. Jaksch \textit{et al.}, Phys. Rev. Lett.  \textbf{81}, 3108 (1998).
\bibitem{PorrasPRL2004}
D. Porras and J. I. Cirac, Phys. Rev. Lett.  \textbf{93}, 263602 (2004).
\bibitem{KasprzakNature2006}
J. Kasprzak and \textit{et al.}, Nature (London) \textbf{443}, 409 (2006).
\bibitem{HDengPRL2006}
H. Deng and \textit{et al.},  Phys. Rev. Lett. \textbf{97}, 146402 (2006)
\end{thebibliography}
\end{document}